\documentclass[twocolumn]{aastex62}

\usepackage{natbib}
\usepackage{amsmath}
\usepackage{braket}
\usepackage[]{hyperref}
\hypersetup{
 colorlinks=true,%
 linkcolor=black,
 citecolor=blue,
}

\begin{document}
\def\vec#1{\mbox{\boldmath $#1$}}
\newcommand{\average}[1]{\ensuremath{\langle#1\rangle} }

\title{\large{\bf Three-dimensional simulation of the fast solar wind driven by compressible magnetohydrodynamic turbulence}}

	\author[0000-0002-7136-8190]{Munehito Shoda}
	\affiliation{National Astronomical Observatory of Japan, National Institute of Natural Sciences, 2-21-1 Osawa, Mitaka, Tokyo, 181-8588, Japan}
	\affiliation{Department of Earth and Planetary Science, The University of Tokyo, 7-3-1 Hongo, Bunkyo, Tokyo, 113-0033, Japan }
	\author[0000-0001-9734-9601]{Takeru Ken Suzuki}
	\affiliation{School of Arts \& Science, The University of Tokyo, 3-8-1 Komaba, Meguro, Tokyo, 153-8902, Japan}
	\author[0000-0003-0204-8385]{Mahboubeh Asgari-Targhi}
	\affiliation{Harvard--Smithsonian Center for Astrophysics, 60 Garden Street MS-15, Cambridge, MA 02138, USA}
	\author[0000-0001-5457-4999]{Takaaki Yokoyama}
	\affiliation{Department of Earth and Planetary Science, The University of Tokyo, 7-3-1 Hongo, Bunkyo, Tokyo, 113-0033, Japan }
	
	\correspondingauthor{Munehito Shoda}
	\email{shoda@eps.s.u-tokyo.ac.jp}

\begin{abstract}

	Using a three-dimensional compressible magnetohydrodynamic (MHD) simulation, 
	we have reproduced the fast solar wind in a direct and self-consistent manner, 
	based on the wave/turbulence driven scenario.
	As a natural consequence of Alfv\'enic perturbations at the coronal base, highly compressional and turbulent fluctuations are generated, leading to heating and acceleration of the solar wind.
	The analysis of power spectra and structure functions reveals that the turbulence is characterized 
	by its imbalanced (in the sense of outward Alfv\'enic fluctuations) and anisotropic nature.
	The density fluctuation originates from the parametric decay instability of outwardly propagating Alfv\'en waves and plays a significant role 
	in the Alfv\'en wave reflection that triggers turbulence. 
	Our conclusion is that the fast solar wind is heated and accelerated by compressible MHD turbulence 
	driven by parametric decay instability and resultant Alfv\'en wave reflection.		
\end{abstract}

\keywords{magnetohydrodynamics (MHD) --
methods: numerical -- Sun: solar wind -- turbulence}

\section{Introduction} \label{sec:introduction}

One of the most important unsolved problems in astrophysics is the driving mechanism of the solar wind.
In addition to the close relation to the coronal heating problem \citep{Parke58}, 
understanding solar wind acceleration is required for stellar rotational evolution \citep[e.g.][]{Brun017} and for space weather forecasting at Earth and at exoplanets \citep[e.g.][]{Garra16}.
It is now widely accepted that the ultimate energy source of the solar wind comes from the surface convection.
Only $0.1-1\%$ of the photospheric energy flux is sufficient to drive the solar wind \citep{Withb77}.
Indeed, some observations confirm the sufficient upward energy transport \citep{DePon07a,McInt11}, whereas we should note that these observations are still controversial \citep[see e.g.][]{Thurg14}.
An unsolved issue regarding solar wind acceleration is the thermalization process, specifically the nature of solar wind turbulence where the solar wind is accelerated.
In-situ observations near $1 {\rm \ au}$ indicate that the turbulent dissipation (cascading) accounts for ongoing heating of the solar wind plasma against adiabatic expansion \citep{Carbo09}.
However, since the plasma condition is very different between the Earth's orbit ($r / R_\odot \approx 200$ where $R_\odot$ denotes the solar radius) and the wind acceleration region ($r / R_\odot \approx 10$), it is risky to simply assume the same situation.
In fact, several observations show that the density fluctuation is large near the Sun \citep[$r/R_\odot \approx 1-10$, see][]{Miyam14,Hahn018}.

In this study, based on the wave/turbulence-driven (WTD) scenario \added{\citep{Hollw86,Ofman98,Ofman04,Suzuk05,Cranm07,Verdi10,Chand11,Cranm12b}}, we perform three-dimensional, compressible MHD simulation of the fast solar wind.
The simulation is conducted in a self-consistent manner; direct calculation of MHD equations enables us to consider the evolutions of the mean field and fluctuation simultaneously.
The compressibility is critical for two reasons.
First, due to the compression of plasma, formation of shock waves is allowed, which can contribute to the heating of the solar wind \added{\citep{Hollw82d,Suzuk05,Matsu14,Shoda18a}}.
Second, the parametric decay instability (PDI) is incorporated.
PDI is an instability of Alfv\'en wave \added{\citep{Galee63,Sagde69,Golds78,Derby78}} and can grow in the wind acceleration region \citep{Suzuk06a,Tener13,Shoda18d,Revil18,Chand18} 
and activate the turbulence by introducing various energy cascading channels \citep[e.g.][]{Shoda18c}.
In fact, \added{some have found that} reduced MHD model cannot account for the solar wind heating without density fluctuation \added{\citep{Perez13,Balle16,Verdi19b}} that is likely to be generated by PDI.
In addition to the compressibility, three-dimensionality is crucial for solving turbulence.
In general, lower-dimensional (1D or 2D) simulations show different behavior compared with 3D ones \citep[e.g.][]{Shoda18c}. 
Therefore, both compressibility and three-dimensionality \replaced{are}{appear to be} crucial for the study of solar wind turbulence.

\section{Method}
We simulate the fast solar wind from the polar region in the solar minimum.
The basic equations are ideal MHD equations with gravity and thermal conduction in the spherical coordinates:
\begin{align}
	\frac{\partial}{\partial t} \rho &+ \nabla \cdot \left( \rho \vec{v} \right) = 0, \\
	\frac{\partial}{\partial t} \left( \rho \vec{v} \right) &+ 
	\nabla \cdot \left( \rho \vec{v} \vec{v} + p_T \hat{\vec{I}} - \frac{\vec{B}\vec{B}}{4\pi} \right) = \rho \vec{g}, \\
	\frac{\partial}{\partial t} \vec{B} &+ \nabla \cdot \left( \vec{v} \vec{B} - \vec{B} \vec{v} \right) = 0, \\
	\frac{\partial}{\partial t}e &+ \nabla \cdot \left[ \left( e + p_T \right) \vec{v} - \frac{\vec{B}}{4 \pi} \left( \vec{v} \cdot \vec{B} \right) + \vec{q}_{\rm cnd} \right] = \rho \vec{g} \cdot \vec{v},
\end{align}
where
\begin{align}
	e = \frac{p}{\gamma - 1} + \frac{1}{2}\rho \vec{v}^2 + \frac{\vec{B}^2}{8 \pi}, \ \ \ p_T = p + \frac{\vec{B}^2}{8 \pi}.
\end{align}
\added{See, for example, \citet{Matsu16} for the notations of variables.}
$\hat{\vec{I}}$ stands for the unit tensor,
$\vec{g}$ and $\vec{q}_{\rm cnd}$ are the gravitational acceleration and thermal conductive flux, respectively.
We employ the adiabatic specific heat ratio of monatomic gas $\gamma = 5/3$.
In solving these equations, the spherical coordinate system is used on the $\theta = \pi/2$ plane that symmetrizes $\theta$ and $\phi$ with respect to $\nabla$ operator as
\begin{align}
	\nabla \approx \vec{e}_r \frac{\partial}{\partial r} 
	+ \vec{e}_\theta \frac{1}{r} \frac{\partial}{\partial \theta}
	+ \vec{e}_\phi \frac{1}{r} \frac{\partial}{\partial \phi},
	\label{eq:nabla}
\end{align}
where $\vec{e}_{r,\theta,\phi}$ denotes the unit vector in each direction.
Due to the small horizontal ($\theta$ and $\phi$) extension of our numerical domain, this approximation yields at most $0.1 \%$ error compared with the usual spherical coordinate.
Note that we replace $r \sin \theta$ with $r$ in Eq. (\ref{eq:nabla}) and the deviation between the two is in the order of $(\theta-\pi/2)^2$ near $\theta=\pi/2$, which yields $10^{-3}$ in our setting.
Since we are simulating polar wind, $\vec{g}$ is given as
\begin{align}
	\vec{g} = - \frac{G M_\odot}{r^2} \vec{e}_r.
\end{align}

The radiative cooling is not considered because we do not solve the atmosphere below the transition region where radiation plays a role.
The thermal conduction instead dominates the energy balance in the corona and solar wind.
We employ a conductive flux that mimics the Spitzer-H\"arm type one \citep{Spitz53} as
\begin{align}
	\vec{q}_{\rm cnd} \cdot \vec{e}_r &= - \kappa_0 \xi(r) T^{5/2} \hat{b}_r^2 \frac{\partial T}{\partial r}, \\
	\vec{q}_{\rm cnd} \cdot \vec{e}_\theta &= - \kappa_0 \xi(r) T^{5/2} f_q \hat{b}_{\theta}^2 \frac{1}{r} \frac{\partial T}{\partial \theta}, 
\end{align}
where $\kappa_0 = 10^{-6}$ in cgs unit, $\xi(r)$ is a quenching due to the large mean free path of electron \citep{Hollw74} and $\hat{b}_{r,\theta,\phi} = B_{r,\theta,\phi}/\left| \vec{B} \right|$.
$\phi$ component of $\vec{q}_{\rm cnd}$ is given in a similar way as $\theta$ component.
The quenching term $\xi(r)$ is given as a function of $r$:
\begin{align}
	\xi(r) = \min \left( 1, \frac{r_{\rm sw}^2}{r^2} \right), \label{eq:conduction_quenching}
\end{align}
where $r_{\rm sw} = 5 R_\odot$ in this study.
\added{Note that Eq. (\ref{eq:conduction_quenching}) yields much stronger conduction quenching compared with observation.
Only a few reduction from Spitzer-H\"arm value is observed in the solar wind \citep{Bale013,Versc19}. 
The $1/r^2$ dependence of $\xi(r)$ is for numerical reason; such $\xi(r)$ makes the thermal diffusivity $\left( \propto \xi(r) T^{5/2} / \rho \right)$ almost constant in $r$.
In the future, we will compute additional models with full Spitzer-Harm conduction and compare the results with the approximate version used here.} 
An additional quenching $f_q = 0.1$ is used to avoid the severe restriction of time step.
Numerical results do not depend on the value of $f_q$ because the conductions in $\theta$ and $\phi$ directions are sufficiently fast to homogenize the temperature on the horizontal plane.
Although the conductive flux in our model is not the same as Spitzer-H\"arm type one, the most important effect of thermal conduction, that is cooling by the radial heat transport, is appropriately solved.
Thus, our simplified conductive flux is appropriate for the solar wind simulation.

\begin{figure*}[t]
\centering
\includegraphics[width=15cm]{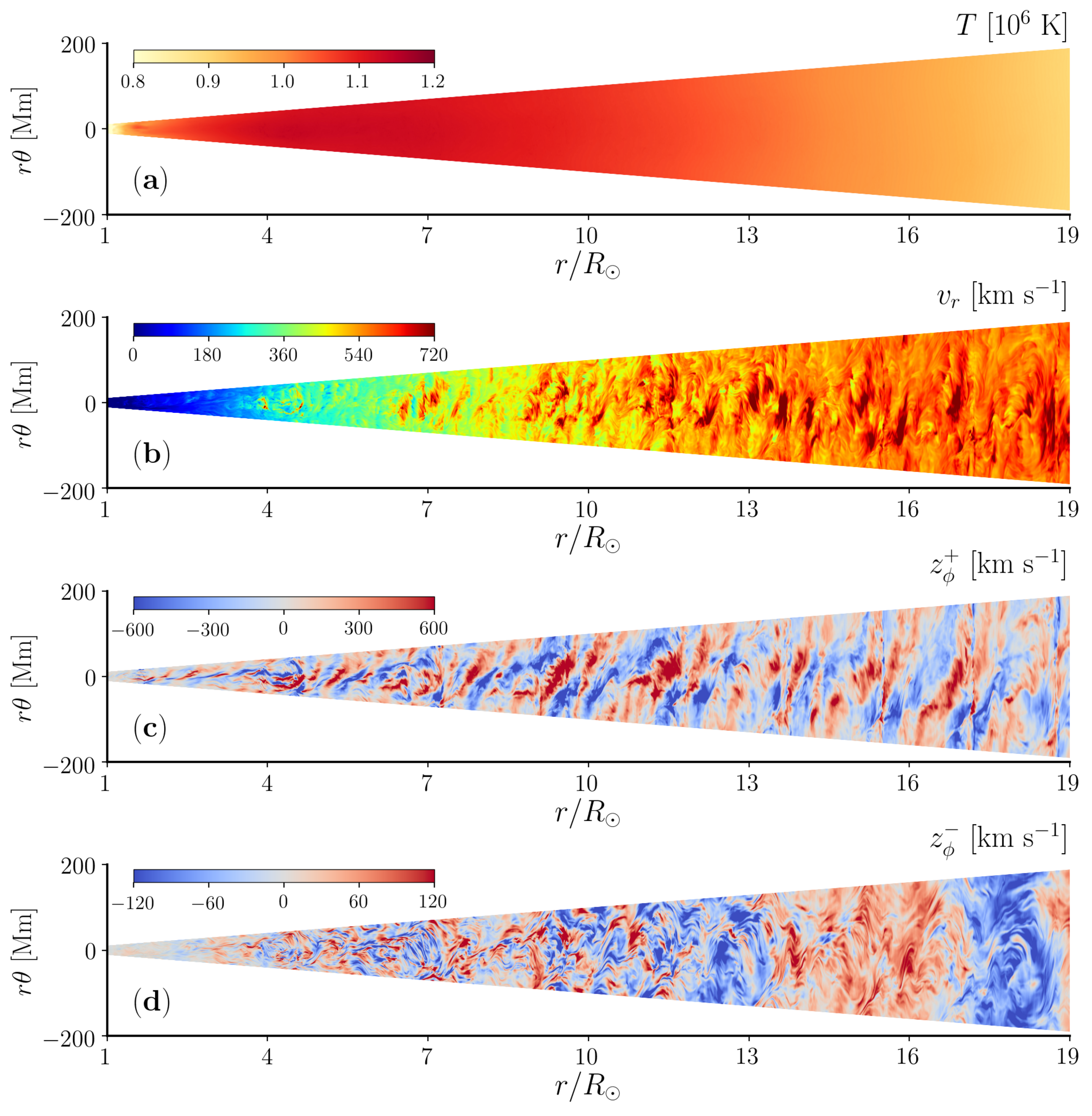}
 \caption{
 	Snapshots on the meridional ($\phi = 0$) plane in the quasi-steady state. Each panel corresponds to (a): temperature, (b): radial velocity, (c): anti-Sunward Els\"asser variable and (d): Sunward Els\"asser variable, respectively.
	An animation of this figure is available in the online journal.
 }     
\vspace{2em} 
\label{meridional_plane}
\end{figure*}
The numerical domain extends from the coronal base ($r=1.02R_\odot$) to $r=20R_\odot$ with horizontal size $20 {\rm \ Mm}$ at the bottom, which yields the range of $\theta$ and $\phi$ as 
\begin{align}
	-1.44 \times 10^{-2} {\rm \ rad} \leq \theta, \ \phi \leq 1.44 \times 10^{-2} {\rm \ rad}.
\end{align}
An additional numerical domain with coarser grids is prepared beyond the top boundary up to $r \approx 10^3 R_{\odot}$, far enough to ensure that no fluctuations reach within the time of simulation.
This method is validated because no physical quantities can propagate back, against the super-Alfv\'enic solar wind, to the numerical domain in the quasi-steady state.
As an initial condition, we impose the isothermal Parker wind with temperature $T=1.1 \times 10^6 {\rm \ K}$ with radially extending magnetic field embedded. 
The bottom boundary is as follows.
The density, temperature and radial magnetic field are fixed to $\rho = 8.5 \times 10^{-16} {\rm  \ g \ cm^{-3}}$, $T = 6 \times 10^5 {\rm \ K}$ and $B_r = 2 {\rm \ G}$, respectively, without any variation in $\theta$ and $\phi$ directions.
\added{We note that the mass-loss rate of the solar wind is basically controlled by the physical condition at the inner boundary. 
In our future study, we take into account the energy transfer from the chromosphere through the transition region, which can determine the mass loss rate in a more self-consistent manner.  \citep{Cranm11,Suzuk13}.}
If we consider the expansion factor of magnetic field, $B_r = 2{\rm \ G}$ is consistent with the source of the fast solar wind \citep{Fujik15},
\added{since in the quasi-steady state of our simulation, the global magnetic field configuration is radial at any given time.}
The radial derivative of $v_r$ is fixed to zero, which allows the supply of mass into the numerical domain.
The inward Els\"asser variables $z_{\theta,\phi}^{-}$ are set to be transmissive at the bottom.
Here the upward ($z_\perp^+$) and downward ($z_\perp^-$) Els\"asser variables are defined as
\begin{align}
	\vec{z}_\perp^\pm = z_{\theta}^\pm \vec{e}_\theta + z_{\phi}^\pm \vec{e}_\phi, \ \ \ \ z_{\theta,\phi}^\pm = v_{\theta,\phi} \mp \frac{B_{\theta,\phi}}{\sqrt{4 \pi \rho}}.
\end{align}
\replaced{The upward Els\"asser variable at the bottom boundary is given with frequency band $10^{-2} - 10^{-3} {\rm \ Hz}$ and amplitude}{The amplitude of the upward Els\"asser variable at the bottom boundary is} $64 {\rm \ km \ s^{-1}}$ corresponding to the observed non-thermal velocity of $32 {\rm \ km \ s^{-1}}$ \citep[e.g.][]{Baner09,Landi09}.
\added{For spatial and temporal profiles of the injected Els\"asser variables, 
we impose $f^{-1}$ spectrum in the range of $10^{-3} {\rm \ Hz} \leq f \leq 10^{-2} {\rm \ Hz}$ for time variation and
$k_\perp^{-2}$ spectrum in the range of $2 \pi / \left( 20 {\rm \ Mm} \right) \leq k_\perp  \leq 6 \pi / \left( 20 {\rm \ Mm} \right)$ for spatial variation.
Note that $20 {\rm \ Mm}$ is the horizontal scale of the simulation domain at the bottom.}
The typical horizontal length scale of the upward Els\"asser variable is fixed to the horizontal size of the simulation domain.
The basic equations are numerically integrated by the combination of 3rd-order SSP Runge--Kutta method \citep{Shu0088} and HLLD Riemann solver \citep{Miyos05} with spatial reconstruction a combination of 2nd-order MUSCL \citep{Leer079} and 5th-order MP5 \citep{Sures97} methods. 
The number of grid points is $(6600,192,192)$ in $(r,\theta,\phi)$ directions, respectively.
The super-time-stepping method is used to solve the thermal conduction \citep{Meyer14}.
To remove the numerically generated finite $\nabla \cdot \vec{B}$, we employ the hyperbolic cleaning method \citep{Dedne02}.

\section{Results \& Discussion}
Figure \ref{meridional_plane} shows the snapshots of temperature (Panel {\bf a}), radial velocity (Panel {\bf b}), anti-Sunward Els\"asser variable (Panel {\bf c}) and Sunward Els\"asser variable (Panel {\bf d}), respectively, on the meridional plane after the quasi-steady state is achieved.
The maximum temperature exceeds $10^6 {\rm \ K}$ and the termination radial velocity approximates $600 {\rm \ km \ s^{-1}}$, 
ensuring the successful reproduction of the fast solar wind.
As a natural consequence of fast thermal conduction, no fine structuring is observed in the temperature map.
The panel of $v_r$ shows that, in addition to the gradual acceleration of the solar wind, ubiquitous local (or discontinuous) enhancements are observed.
According to the previous 1D simulations \citep{Suzuk05,Shoda18a}, these fluctuations are large-amplitude slow mode waves that can at least partially contribute to the heating of the solar wind.

Panels {\bf c} and {\bf d} in Figure \ref{meridional_plane} show the evolution of waves and turbulence.
Note that $z_\phi^+$ and $z_\phi^-$ correspond to anti-Sunward and Sunward Alfv\'en wave characteristics in the linear regime.
$z_\phi^+$ maintains the coherent structure in the entire simulation domain while $z_\phi^-$ shows an evidence of strong turbulent distortion.
\begin{figure}[t]
\centering
\includegraphics[width=7.5cm]{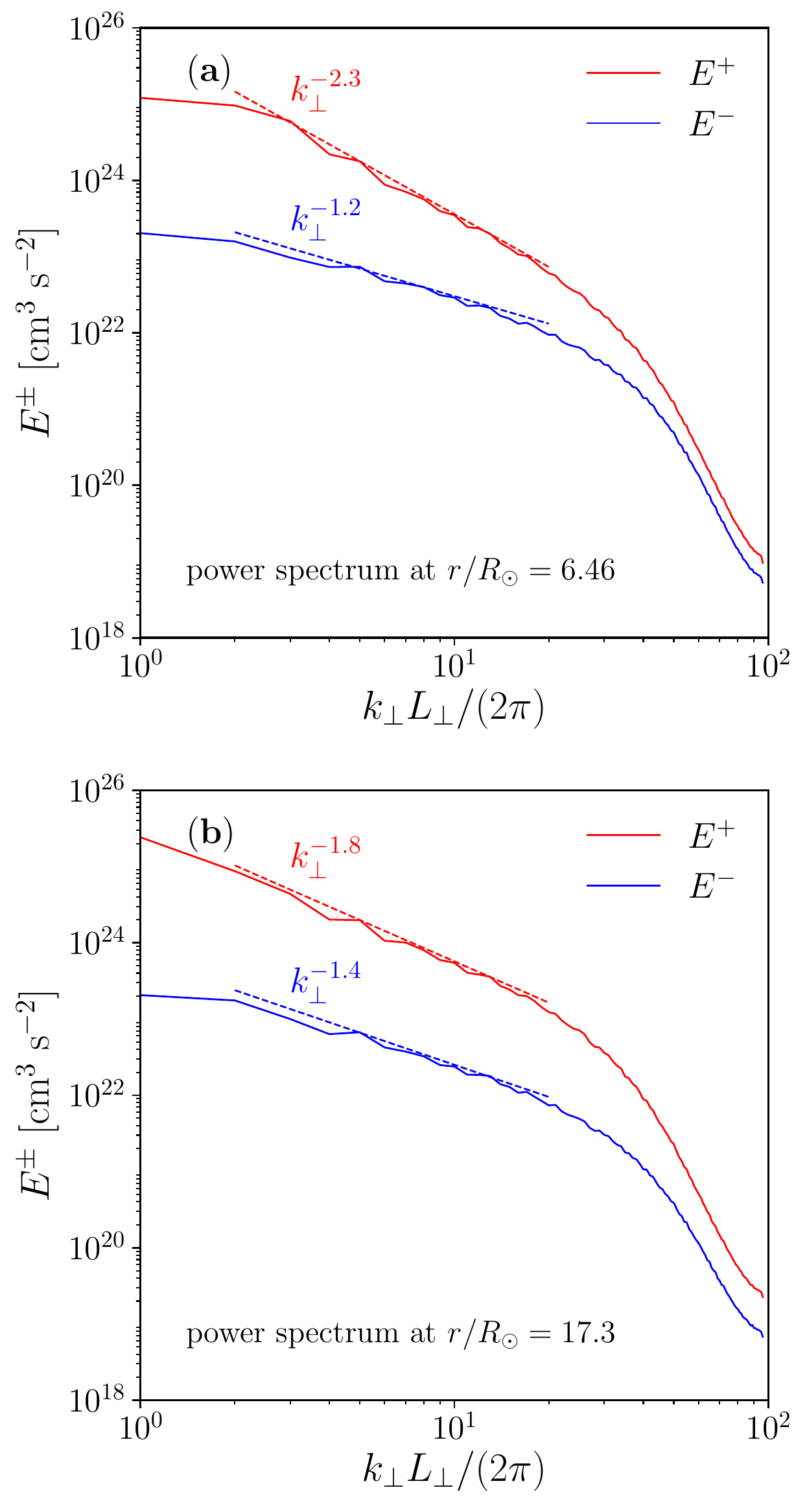}
 \caption{
 	Omnidirectional Els\"asser power spectra with respect to perpendicular wave number.
	The top and bottom panels correspond to the radial distances of $r/R_\odot = 6.46$ and $r/R_\odot = 17.3$, respectively.
	Solid lines indicate the spectra of anti-Sunward ($E^+$, red) and Sunward ($E^-$, blue) Els\"asser variables, respectively.
	Power-law-fitted lines in the inertial range are shown by the dashed lines.
 }     
\label{wavenumber_spectrum}
\vspace{1em}
\end{figure}
It is evident, especially in $4 \lesssim r/R_\odot \lesssim 10$, that $z_\phi^-$ has much finer transverse structure than $z_\phi^+$.
This feature is more quantitatively observed in the Els\"asser power spectra with respect to perpendicular wave number defined as
\begin{align}
	\int E^\pm \left( k_\perp \right) dk_\perp = \frac{1}{L_\perp^2} \int r^2 d\theta d\phi \ \vec{z}_\perp^2 \left(\theta,\phi \right),
\end{align}
where $k_\perp$ is the wave number perpendicular to the mean field direction ($r$ axis) and $L_\perp$ denotes the horizontal extension of the simulation domain at $r$.
Note that $E^\pm \left( k_\perp \right)$ reflects the spatial structures of anti-Sunward and Sunward Els\"asser variables perpendicular to the mean field.
Solid lines in Figure \ref{wavenumber_spectrum} show $E^\pm \left( k_\perp \right)$ calculated at $r/R_\odot = 6.46$ (Panel {\bf a}) and $r/R_\odot = 17.3$ (Panel {\bf b}).
Also shown by dashed lines correspond to the power-law fitting in the inertial range ($2 \leq k_\perp L_\perp / (2 \pi) \leq 20$).
We observe different inertial-range power indices between $E^+$ and $E^-$ for both locations.
Specifically, the anti-Sunward component has flatter (harder) power spectrum than that of Sunward one.
This spectral behavior is consistent with structure difference between $z_\phi^+$ and $z_\phi^-$ observed in Figure \ref{meridional_plane},
because the flatter power spectrum is associated with finer structures.
\added{Note that similar power spectra of $E^\pm$ are found in imbalanced incompressible MHD turbulence \citep{Chand08}.}
Another interesting feature is that, as $r$ gets larger, the power indices of both $E^+$ and $E^-$ approach the Kolmogorov's index $-5/3$, which is observed in the magnetic power spectrum in the solar wind \citep[e.g.][]{Bruno13}.
\added{See below for additional discussion of the turbulence physics.}

A brief description of the spectral difference is given as follows.
In the regime of reduced MHD, neglecting the inhomogeneity of the background, the evolution of Alfv\'en waves is described as follows \citep[e.g.][]{Pries14}:
\begin{align}
	\frac{\partial}{\partial t} \vec{z}^\pm + \left( \vec{z}^\mp \cdot \nabla \right) \vec{z}^\pm = 0, \ \ \vec{z}^\pm = \vec{v} \mp \vec{B} /\sqrt{4 \pi \rho}.
	\label{eq:elsasser}
\end{align}
Thus, the nonlinear wave-wave interaction is invoked by the collision of counter-propagating waves.
Note, however, that, in the presence of background inhomogeneity, this is not the case because Els\"asser variables are no longer pure characteristics of Alfv\'en waves \citep[anomalous components, see][]{Velli89,Velli93,Perez13}.
Eq. (\ref{eq:elsasser}) shows that the energy cascading timescale of $z^\pm$ is determined by $z^\mp$.
Specifically when $z^+ > z^-$, the cascading of $z^-$ proceeds faster than $z^+$, leading to structure difference \added{\citep[see e.g.][]{Chand15}}.
In terms of relaxation process, this process is called dynamical alignment \citep[e.g.][]{Biska03}, in which the minor component of $z^\pm$ decays faster than the major one.

More quantitative explanation of the spectral imbalance is also given both numerically and analytically.
The theory and simulation of the incompressible MHD turbulence show \citep{Boldy09}
\begin{align}
	E^\pm \propto k_\perp^{-2 \pm \alpha},
\end{align}
while the strong turbulence (EDQNM) theory predicts \citep{Grapp83}
\begin{align}
	E^\pm \propto k_\perp^{-3/2 \pm \tilde{\alpha}},
\end{align}
where $\alpha$ and $\tilde{\alpha}$ depend on the degree of imbalance $E^+/E^-$.
Compressible MHD turbulence also possibly exhibits the similar spectral difference \citep{Perez12}.
Though not perfectly, our results are at least qualitatively consistent with these predictions.
The summation of power indices shifts from $-3.5$ ($r/R_\odot = 6.46$) to $-3.2$ ($r/R_\odot = 17.3$), suggesting the weak-to-strong transition of turbulence. 

\begin{figure}[t]
\centering
\includegraphics[width=7.5cm]{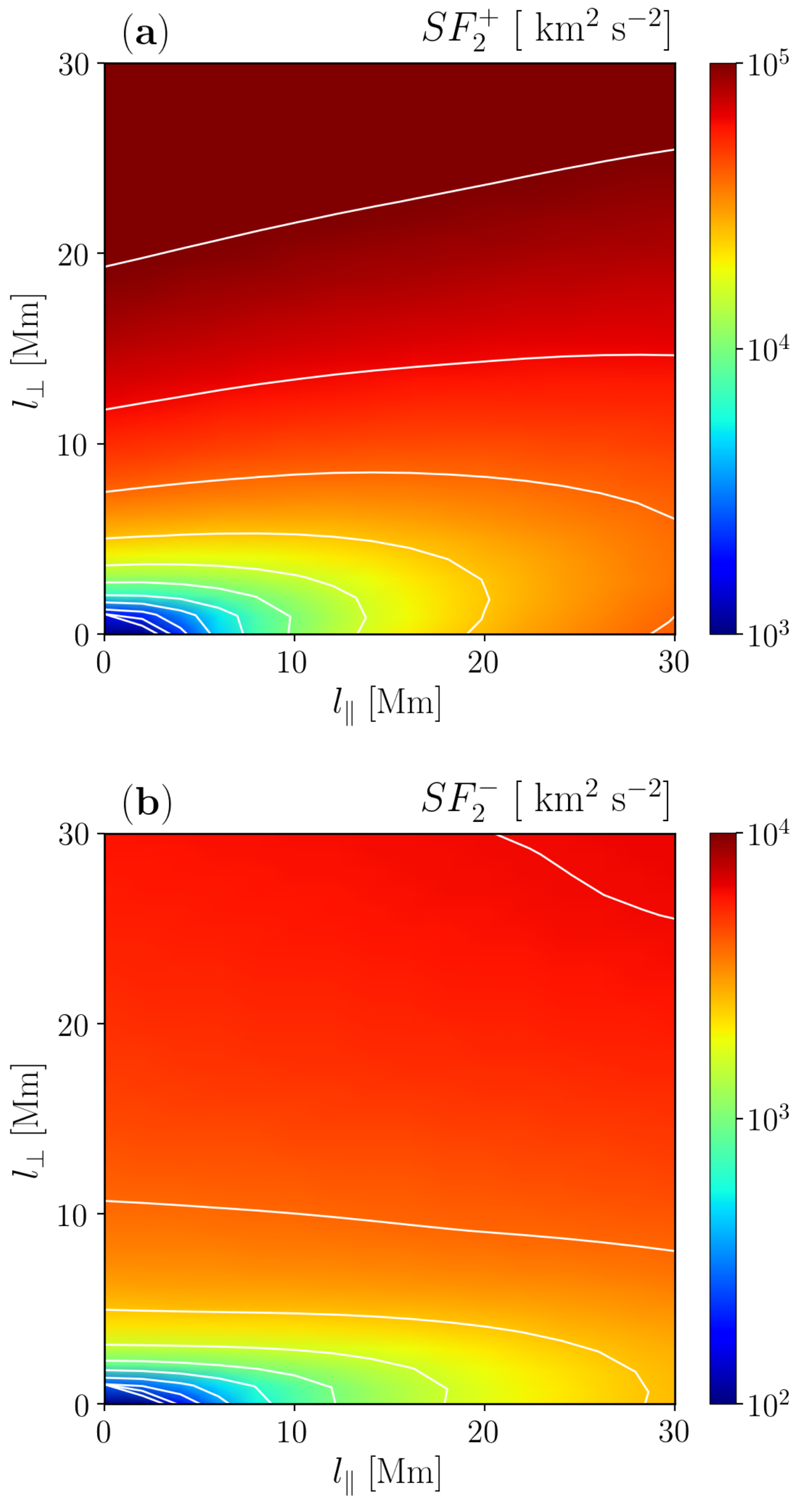}
 \caption{
 	2nd-order structure function of anti-Sunward ($SF_2^+$, top) and Sunward ($SF_2^-$, bottom) Els\"asser variables measured 
	in the vicinity of $r/R_\odot = 10$.
	Embedded white lines represent the contour lines.
 }     
 \label{structure_function}
\end{figure}
The anisotropy is another factor that characterizes the structure of turbulence.
In the presence of mean magnetic field, the structure of turbulence is expected to be anisotropic \citep[e.g.][]{Goldr95}.
To see the degree of anisotropy, we often use the 2nd-order structure function \citep{Cho0003,Verdi15} defined as 
\begin{align}
	SF_2^\pm \left( l_\parallel, l_\perp \right) &= \langle \left| \vec{z}_\perp^\pm \left( \vec{x} + l_\parallel \vec{e}_r + l_\perp \vec{e}_\theta \right) -  \vec{z}_\perp^\pm \left( \vec{x} \right) \right|^2 \rangle
\end{align}
where the bracket denotes the averaging operator over $\theta$, $\phi$ and $t$.
Here the structure function is defined based on two assumptions; the turbulence is anisotropic with respect to $r$ axis and isotropic in the $\theta \phi$ plane.
The first assumption is justified since the mean magnetic field is perfectly aligned with $r$ axis.
In some simulations of the solar wind, the second assumption is violated because the wind expansion introduces another form of anisotropy \citep{Dong014}.
The flow direction of the solar wind is therefore an additional anisotropy axis that forms 3D anisotropy of the wind turbulence \citep{Verdi18}.
In our calculation, because we simulate the wind from the polar region, the flow direction is aligned with the mean field.
The definition of the structure function is justified.

Figure \ref{structure_function} shows $SF_2^\pm$ measured at $r/R_\odot =10$.
Both $SF_2^+$ and $SF_2^-$ rapidly increase in $l_\perp$ direction, showing that field-aligned structures are generated preferentially in the solar wind.
This is consistent with previous works \citep[e.g.][]{Cho0003,Shoda18c}.
A difference of anisotropy is also observed; the minor component ($SF_2^-$) shows larger degree of anisotropy than the major component ($SF_2^+$).
This is consistent with the result of phenomenological study of Alfv\'en wave turbulence \citep{Beres08}.
We also note that the fluctuations in the solar wind show a similar behavior \citep{Wicks11}.

\begin{figure*}[t]
\centering
 \includegraphics[width=15cm]{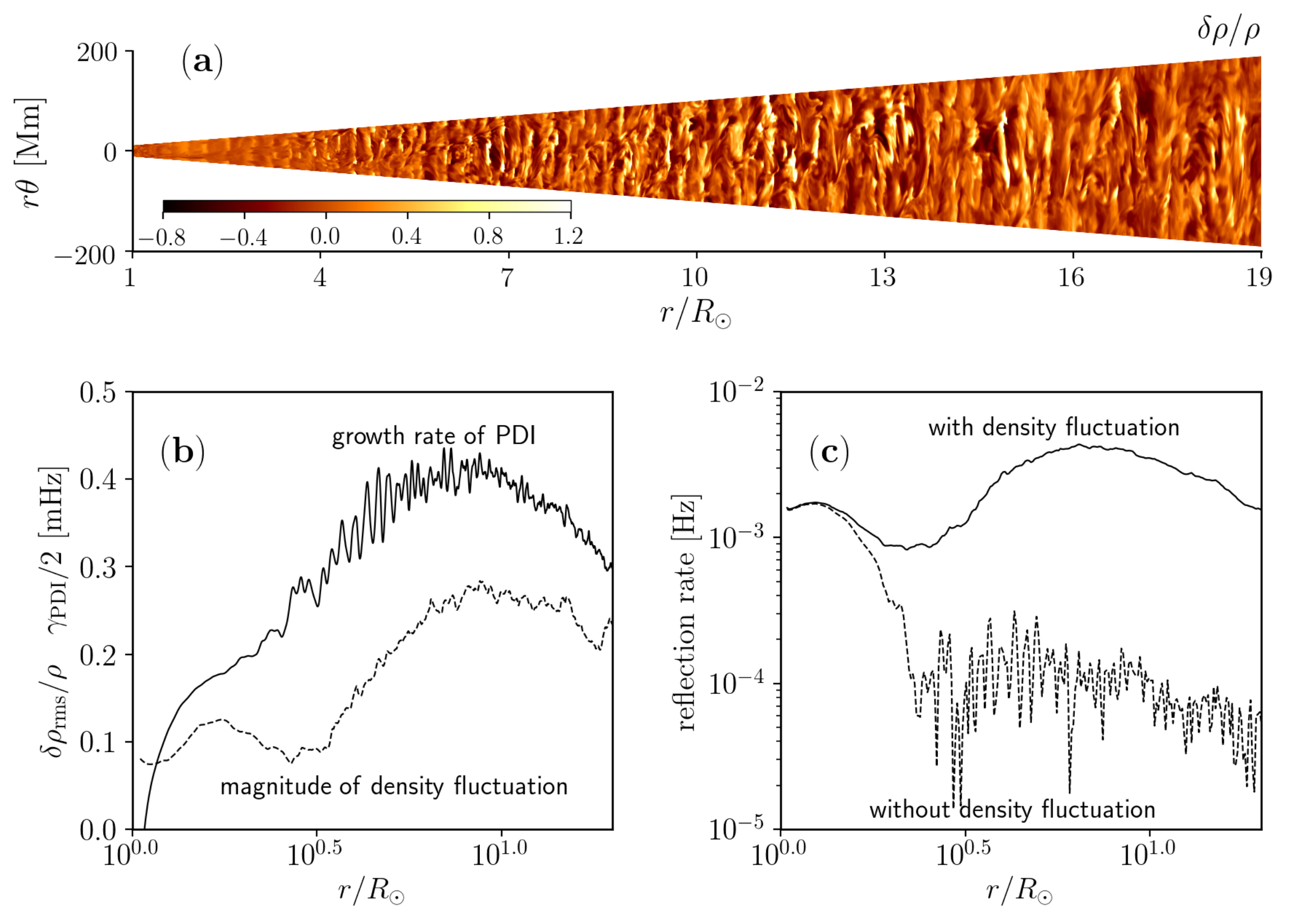}
 \caption{
 	Panel {\bf a}: a snapshot of fractional density fluctuation $\delta \rho / \rho$ in the meridional plane.
	Panel {\bf b}: growth rate of parametric decay instability ($\gamma_{\rm PDI}/2$, solid line) in unit of ${\rm mHz}$ and the magnitude of density fluctuation ($\delta \rho / \rho$, dashed line) versus radial distance.
	Panel {\bf c}: reflection rates of anti-Sunward Alfv\'en waves $\omega_{\rm ref}$ with and without density fluctuations.
	An animation of Panel {\bf a} is available in the online journal.
 }     
 \label{density_fluctuation}
 \vspace{1em}
\end{figure*}
The results and discussion above show that the structure of solar wind turbulence is consistent with that of incompressible MHD (Alfv\'en-wave) turbulence.
This does not mean that the compressibility is ignorable in the solar wind.
In fact, the density fluctuation \replaced{is}{appears to be} required to account for the heating rate in the solar wind \citep{Balle16}.
Specifically, the density fluctuations act as reflectors of anti-Sunward Alfv\'en waves, playing an indirect but critical role in the onset of turbulence.
Figure \ref{density_fluctuation} shows the distribution, origin and role of density fluctuation.
Panel (a) shows the turbulent and discontinuous structure of fractional density fluctuation $\delta \rho / \rho$ on the meridional plane.
Note that the magnitude of density fluctuation is, at least locally, as large as the mean density.
Panel (b) displays the magnitude of \added{root-mean-squared fractional} density fluctuation, \added{$\delta \rho_{\rm rms} / \rho$, where
\begin{align}
	\delta \rho_{\rm rms}= \sqrt{\langle \rho^2 \rangle -  \langle \rho \rangle^2},
\end{align}
}
and the growth rate of parametric decay instability (PDI) of Alfv\'en waves, $\gamma_{\rm PDI}$, in unit of ${\rm \ mHz}$.
In the accelerating, expanding solar wind, $\gamma_{\rm PDI}$ is given as \citep{Tener13,Shoda18d}
\begin{align}
	\gamma_{\rm PDI} = \tilde{\gamma}_{\rm GD} 2 \pi f_0 - \gamma_{\rm acc} - \gamma_{\rm exp},
\end{align}
where $\tilde{\gamma}_{\rm GD}$ is the normalized growth rate \citep{Golds78,Derby78} in the homogeneous system, $f_0 = 10^{-3} {\rm \ Hz}$ is the lowest frequency of the injected Alfv\'en waves, and $\gamma_{\rm acc}$ and $\gamma_{\rm exp}$ are the suppression of density fluctuation by the wind acceleration and expansion \citep[see][]{Shoda18d}.
Note that we show $\gamma_{\rm PDI}/2$ rather than $\gamma_{\rm PDI}$ for better visualization.
A clear spatial correlation between $\gamma_{\rm PDI}$ and $\delta \rho / \rho$ shows that the density fluctuation originates in PDI.
Thus \deleted{it is probable that} the density fluctuation comes from the PDI.
Panel (c) shows how the density fluctuation affects the Alfv\'en wave propagation by displaying the reflection rate of anti-Sunward Alfv\'en waves given as \citep{Heine80}
\begin{align}
	\omega_{\rm ref} = \left( v_r + v_{A,r} \right) \frac{\partial}{\partial r} \ln \left( r \rho^{1/4} \right),
	\label{eq:reflection}
\end{align}
where $v_{A,r} = B_r / \sqrt{4 \pi \rho}$.
To see the role of density fluctuation, we compare $\omega_{\rm ref}$ calculated in two ways.
First, we calculate $\omega_{\rm ref}$ at each given time and space and average it over time and $\theta \phi$ plane.
Second, we first average $v_r$, $v_{A,r}$ and $\rho$ in time and horizontal space, and calculate $\omega_{\rm ref}$ using Eq. (\ref{eq:reflection}).
The former and latter correspond to the reflection rates with and without density fluctuations, because any fluctuations are smoothed out by averaging the background values ($v_r$, $v_{A,r}$, $\rho$).
The comparison of these two values are shown in Panel (c).
The density fluctuation enhances the reflection rate by a factor of 10 or larger, and therefore the density fluctuation plays a dominant role in the Alfv\'en wave reflection that triggers turbulence.
To summarize, we have shown that the density fluctuation, excited by the PDI, plays a crucial role in the turbulence trigger, and therefore, the compressibility is far from negligible in the solar wind acceleration.

\section{Conclusion}
Our three-dimensional MHD simulation reproduces the fast solar wind as a natural consequence of Alfv\'en-wave injection from the coronal base, 
thus supporting the wave/turbulence-driven model of the fast solar wind.
The turbulence is characterized by imbalance (Figure \ref{wavenumber_spectrum}), anisotropy (Figure \ref{structure_function}), and compressibility (Figure \ref{density_fluctuation}).
The structure of turbulence is well described by the incompressible or reduced MHD turbulence in which the compressibility is ignored.
To discuss the turbulent dissipation and heating, however, compressibility plays a crucial role because the wave reflection, the source of Alfv\'en wave turbulence, is driven dominantly by density fluctuations excited by the parametric decay instability of Alfv\'en waves.

We acknowledge the anonymous referee for a number of helpful and constructive comments.
We would also like to thank Marco Velli and Takuma Matsumoto for many useful discussions and comments.
Numerical computations were carried out on Cray XC50 at Center for Computational Astrophysics, National Astronomical Observatory of Japan.
M. Shoda is supported by Grant-in-Aid for Japan Society for the Promotion of Science (JSPS) Fellows and 
by the NINS program for cross-disciplinary study (Grant Nos. 01321802 and 01311904) 
on Turbulence, Transport, and Heating Dynamics in Laboratory and Solar/Astrophysical Plasmas: "SoLaBo-X".
T.K. Suzuki is supported in part by Grants-in-Aid for Scientific Research from the MEXT of Japan, 17H01105.
M. Asgari-Targhi was supported under contract NNM07AB07C from NASA to the
Smithsonian Astrophysical Observatory (SAO) and contract SP02H1701R from Lockheed
Martin Space and Astrophysics Laboratory (LMSAL) to SAO.
T. Yokoyama is supported by JSPS KAKENHI Grant Number 15H03640.
	
\bibliographystyle{apj}

\begin{thebibliography}{68}
\expandafter\ifx\csname natexlab\endcsname\relax\def\natexlab#1{#1}\fi

\bibitem[{{Bale} {et~al.}(2013){Bale}, {Pulupa}, {Salem}, {Chen}, \&
  {Quataert}}]{Bale013}
{Bale}, S.~D., {Pulupa}, M., {Salem}, C., {Chen}, C.~H.~K., \& {Quataert}, E.
  2013, \apj, 769, L22

\bibitem[{{Banerjee} {et~al.}(2009){Banerjee}, {P{\'e}rez-Su{\'a}rez}, \&
  {Doyle}}]{Baner09}
{Banerjee}, D., {P{\'e}rez-Su{\'a}rez}, D., \& {Doyle}, J.~G. 2009, \aap, 501,
  L15

\bibitem[{{Beresnyak} \& {Lazarian}(2008)}]{Beres08}
{Beresnyak}, A., \& {Lazarian}, A. 2008, \apj, 682, 1070

\bibitem[{{Biskamp}(2003)}]{Biska03}
{Biskamp}, D. 2003, {Magnetohydrodynamic Turbulence (Cambridge: Cambridge University Press)}

\bibitem[{{Boldyrev} \& {Perez}(2009)}]{Boldy09}
{Boldyrev}, S., \& {Perez}, J.~C. 2009, Physical Review Letters, 103, 225001

\bibitem[{{Brun} \& {Browning}(2017)}]{Brun017}
{Brun}, A.~S., \& {Browning}, M.~K. 2017, Living Reviews in Solar Physics, 14,
  4

\bibitem[{{Bruno} \& {Carbone}(2013)}]{Bruno13}
{Bruno}, R., \& {Carbone}, V. 2013, Living Reviews in Solar Physics, 10, 2

\bibitem[{{Carbone} {et~al.}(2009){Carbone}, {Marino}, {Sorriso-Valvo},
  {Noullez}, \& {Bruno}}]{Carbo09}
{Carbone}, V., {Marino}, R., {Sorriso-Valvo}, L., {Noullez}, A., \& {Bruno}, R.
  2009, Physical Review Letters, 103, 061102

\bibitem[{{Chandran}(2008)}]{Chand08}
{Chandran}, B.~D.~G. 2008, \apj, 685, 646

\bibitem[{{Chandran}(2018)}]{Chand18}
---. 2018, Journal of Plasma Physics, 84, 905840106

\bibitem[{{Chandran} {et~al.}(2011){Chandran}, {Dennis}, {Quataert}, \&
  {Bale}}]{Chand11}
{Chandran}, B.~D.~G., {Dennis}, T.~J., {Quataert}, E., \& {Bale}, S.~D. 2011,
  \apj, 743, 197

\bibitem[{{Chandran} {et~al.}(2015){Chandran}, {Schekochihin}, \&
  {Mallet}}]{Chand15}
{Chandran}, B.~D.~G., {Schekochihin}, A.~A., \& {Mallet}, A. 2015, \apj, 807,
  39

\bibitem[{{Cho} \& {Lazarian}(2003)}]{Cho0003}
{Cho}, J., \& {Lazarian}, A. 2003, \mnras, 345, 325

\bibitem[{{Cranmer}(2012)}]{Cranm12b}
{Cranmer}, S.~R. 2012, \ssr, 172, 145

\bibitem[{{Cranmer} \& {Saar}(2011)}]{Cranm11}
{Cranmer}, S.~R., \& {Saar}, S.~H. 2011, \apj, 741, 54

\bibitem[{{Cranmer} {et~al.}(2007){Cranmer}, {van Ballegooijen}, \&
  {Edgar}}]{Cranm07}
{Cranmer}, S.~R., {van Ballegooijen}, A.~A., \& {Edgar}, R.~J. 2007, \apjs,
  171, 520

\bibitem[{{De Pontieu} {et~al.}(2007){De Pontieu}, {McIntosh}, {Carlsson},
  {Hansteen}, {Tarbell}, {Schrijver}, {Title}, {Shine}, {Tsuneta}, {Katsukawa},
  {Ichimoto}, {Suematsu}, {Shimizu}, \& {Nagata}}]{DePon07a}
{De Pontieu}, B., {et~al.} 2007, Science, 318, 1574

\bibitem[{{Dedner} {et~al.}(2002){Dedner}, {Kemm}, {Kr{\"o}ner}, {Munz},
  {Schnitzer}, \& {Wesenberg}}]{Dedne02}
{Dedner}, A., {Kemm}, F., {Kr{\"o}ner}, D., {Munz}, C.-D., {Schnitzer}, T., \&
  {Wesenberg}, M. 2002, Journal of Computational Physics, 175, 645

\bibitem[{{Derby}(1978)}]{Derby78}
{Derby}, Jr., N.~F. 1978, \apj, 224, 1013

\bibitem[{{Dong} {et~al.}(2014){Dong}, {Verdini}, \& {Grappin}}]{Dong014}
{Dong}, Y., {Verdini}, A., \& {Grappin}, R. 2014, \apj, 793, 118

\bibitem[{{Fujiki} {et~al.}(2015){Fujiki}, {Tokumaru}, {Iju}, {Hakamada}, \&
  {Kojima}}]{Fujik15}
{Fujiki}, K., {Tokumaru}, M., {Iju}, T., {Hakamada}, K., \& {Kojima}, M. 2015,
  \solphys, 290, 2491

\bibitem[{{Galeev} \& {Oraevskii}(1963)}]{Galee63}
{Galeev}, A.~A., \& {Oraevskii}, V.~N. 1963, Soviet Physics Doklady, 7, 988

\bibitem[{{Garraffo} {et~al.}(2016){Garraffo}, {Drake}, \& {Cohen}}]{Garra16}
{Garraffo}, C., {Drake}, J.~J., \& {Cohen}, O. 2016, \apjl, 833, L4

\bibitem[{{Goldreich} \& {Sridhar}(1995)}]{Goldr95}
{Goldreich}, P., \& {Sridhar}, S. 1995, \apj, 438, 763

\bibitem[{{Goldstein}(1978)}]{Golds78}
{Goldstein}, M.~L. 1978, \apj, 219, 700

\bibitem[{{Grappin} {et~al.}(1983){Grappin}, {Leorat}, \& {Pouquet}}]{Grapp83}
{Grappin}, R., {Leorat}, J., \& {Pouquet}, A. 1983, \aap, 126, 51

\bibitem[{{Hahn} {et~al.}(2018){Hahn}, {D'Huys}, \& {Savin}}]{Hahn018}
{Hahn}, M., {D'Huys}, E., \& {Savin}, D.~W. 2018, \apj, 860, 34

\bibitem[{{Heinemann} \& {Olbert}(1980)}]{Heine80}
{Heinemann}, M., \& {Olbert}, S. 1980, \jgr, 85, 1311

\bibitem[{{Hollweg}(1974)}]{Hollw74}
{Hollweg}, J.~V. 1974, \jgr, 79, 3845

\bibitem[{{Hollweg}(1982)}]{Hollw82d}
---. 1982, \apj, 254, 806

\bibitem[{{Hollweg}(1986)}]{Hollw86}
---. 1986, \jgr, 91, 4111

\bibitem[{{Landi} \& {Cranmer}(2009)}]{Landi09}
{Landi}, E., \& {Cranmer}, S.~R. 2009, \apj, 691, 794

\bibitem[{{Matsumoto} \& {Suzuki}(2014)}]{Matsu14}
{Matsumoto}, T., \& {Suzuki}, T.~K. 2014, \mnras, 440, 971

\bibitem[{{Matsumoto} {et~al.}(2016){Matsumoto}, {Asahina}, {Kudoh},
  {Kawashima}, {Matsumoto}, {Takahashi}, {Minoshima}, {Zenitani}, {Miyoshi}, \&
  {Matsumoto}}]{Matsu16}
{Matsumoto}, Y., {et~al.} 2016, arXiv e-prints, arXiv:1611.01775

\bibitem[{{McIntosh} {et~al.}(2011){McIntosh}, {de Pontieu}, {Carlsson},
  {Hansteen}, {Boerner}, \& {Goossens}}]{McInt11}
{McIntosh}, S.~W., {de Pontieu}, B., {Carlsson}, M., {Hansteen}, V., {Boerner},
  P., \& {Goossens}, M. 2011, \nat, 475, 477

\bibitem[{{Meyer} {et~al.}(2014){Meyer}, {Balsara}, \& {Aslam}}]{Meyer14}
{Meyer}, C.~D., {Balsara}, D.~S., \& {Aslam}, T.~D. 2014, Journal of
  Computational Physics, 257, 594

\bibitem[{{Miyamoto} {et~al.}(2014){Miyamoto}, {Imamura}, {Tokumaru}, {Ando},
  {Isobe}, {Asai}, {Shiota}, {Toda}, {H{\"a}usler}, {P{\"a}tzold}, {Nabatov},
  \& {Nakamura}}]{Miyam14}
{Miyamoto}, M., {et~al.} 2014, \apj, 797, 51

\bibitem[{{Miyoshi} \& {Kusano}(2005)}]{Miyos05}
{Miyoshi}, T., \& {Kusano}, K. 2005, Journal of Computational Physics, 208, 315

\bibitem[{{Ofman}(2004)}]{Ofman04}
{Ofman}, L. 2004, Journal of Geophysical Research (Space Physics), 109, A07102

\bibitem[{{Ofman} \& {Davila}(1998)}]{Ofman98}
{Ofman}, L., \& {Davila}, J.~M. 1998, \jgr, 103, 23677

\bibitem[{{Parker}(1958)}]{Parke58}
{Parker}, E.~N. 1958, \apj, 128, 664

\bibitem[{{Perez} \& {Chandran}(2013)}]{Perez13}
{Perez}, J.~C., \& {Chandran}, B.~D.~G. 2013, \apj, 776, 124

\bibitem[{{Perez} {et~al.}(2012){Perez}, {Mason}, {Boldyrev}, \&
  {Cattaneo}}]{Perez12}
{Perez}, J.~C., {Mason}, J., {Boldyrev}, S., \& {Cattaneo}, F. 2012, Physical
  Review X, 2, 041005

\bibitem[{{Priest}(2014)}]{Pries14}
{Priest}, E. 2014, {Magnetohydrodynamics of the Sun (Cambridge: Cambride University Press)}

\bibitem[{{R{\'e}ville} {et~al.}(2018){R{\'e}ville}, {Tenerani}, \&
  {Velli}}]{Revil18}
{R{\'e}ville}, V., {Tenerani}, A., \& {Velli}, M. 2018, \apj, 866, 38

\bibitem[{{Sagdeev} \& {Galeev}(1969)}]{Sagde69}
{Sagdeev}, R.~Z., \& {Galeev}, A.~A. 1969, {Nonlinear Plasma Theory (New York: Benjamin)}

\bibitem[{{Shoda} \& {Yokoyama}(2018)}]{Shoda18c}
{Shoda}, M., \& {Yokoyama}, T. 2018, \apjl, 859, L17

\bibitem[{{Shoda} {et~al.}(2018{\natexlab{a}}){Shoda}, {Yokoyama}, \&
  {Suzuki}}]{Shoda18a}
{Shoda}, M., {Yokoyama}, T., \& {Suzuki}, T.~K. 2018{\natexlab{a}}, \apj, 853,
  190

\bibitem[{{Shoda} {et~al.}(2018{\natexlab{b}}){Shoda}, {Yokoyama}, \&
  {Suzuki}}]{Shoda18d}
---. 2018{\natexlab{b}}, \apj, 860, 17

\bibitem[{{Shu} \& {Osher}(1988)}]{Shu0088}
{Shu}, C.-W., \& {Osher}, S. 1988, Journal of Computational Physics, 77, 439

\bibitem[{{Spitzer} \& {H{\"a}rm}(1953)}]{Spitz53}
{Spitzer}, L., \& {H{\"a}rm}, R. 1953, Physical Review, 89, 977

\bibitem[{{Suresh} \& {Huynh}(1997)}]{Sures97}
{Suresh}, A., \& {Huynh}, H.~T. 1997, Journal of Computational Physics, 136, 83

\bibitem[{{Suzuki} {et~al.}(2013){Suzuki}, {Imada}, {Kataoka}, {Kato},
  {Matsumoto}, {Miyahara}, \& {Tsuneta}}]{Suzuk13}
{Suzuki}, T.~K., {Imada}, S., {Kataoka}, R., {Kato}, Y., {Matsumoto}, T.,
  {Miyahara}, H., \& {Tsuneta}, S. 2013, \pasj, 65, 98

\bibitem[{{Suzuki} \& {Inutsuka}(2005)}]{Suzuk05}
{Suzuki}, T.~K., \& {Inutsuka}, S.-i. 2005, \apjl, 632, L49

\bibitem[{{Suzuki} \& {Inutsuka}(2006)}]{Suzuk06a}
{Suzuki}, T.~K., \& {Inutsuka}, S.-I. 2006, Journal of Geophysical Research
  (Space Physics), 111, 6101

\bibitem[{{Tenerani} \& {Velli}(2013)}]{Tener13}
{Tenerani}, A., \& {Velli}, M. 2013, Journal of Geophysical Research (Space
  Physics), 118, 7507

\bibitem[{{Thurgood} {et~al.}(2014){Thurgood}, {Morton}, \&
  {McLaughlin}}]{Thurg14}
{Thurgood}, J.~O., {Morton}, R.~J., \& {McLaughlin}, J.~A. 2014, \apjl, 790, L2

\bibitem[{{van Ballegooijen} \& {Asgari-Targhi}(2016)}]{Balle16}
{van Ballegooijen}, A.~A., \& {Asgari-Targhi}, M. 2016, \apj, 821, 106

\bibitem[{{van Leer}(1979)}]{Leer079}
{van Leer}, B. 1979, Journal of Computational Physics, 32, 101

\bibitem[{{Velli}(1993)}]{Velli93}
{Velli}, M. 1993, \aap, 270, 304

\bibitem[{{Velli} {et~al.}(1989){Velli}, {Grappin}, \& {Mangeney}}]{Velli89}
{Velli}, M., {Grappin}, R., \& {Mangeney}, A. 1989, Physical Review Letters,
  63, 1807

\bibitem[{{Verdini} {et~al.}(2018){Verdini}, {Grappin}, {Alexandrova}, \&
  {Lion}}]{Verdi18}
{Verdini}, A., {Grappin}, R., {Alexandrova}, O., \& {Lion}, S. 2018, \apj, 853,
  85

\bibitem[{{Verdini} {et~al.}(2015){Verdini}, {Grappin}, {Hellinger}, {Landi},
  \& {M{\"u}ller}}]{Verdi15}
{Verdini}, A., {Grappin}, R., {Hellinger}, P., {Landi}, S., \& {M{\"u}ller},
  W.~C. 2015, \apj, 804, 119

\bibitem[{{Verdini} {et~al.}(2019){Verdini}, {Grappin}, \&
  {Montagud-Camps}}]{Verdi19b}
{Verdini}, A., {Grappin}, R., \& {Montagud-Camps}, V. 2019, \solphys, 294, 65

\bibitem[{{Verdini} {et~al.}(2010){Verdini}, {Velli}, {Matthaeus}, {Oughton},
  \& {Dmitruk}}]{Verdi10}
{Verdini}, A., {Velli}, M., {Matthaeus}, W.~H., {Oughton}, S., \& {Dmitruk}, P.
  2010, \apjl, 708, L116

\bibitem[{{Verscharen} {et~al.}(2019){Verscharen}, {Klein}, \&
  {Maruca}}]{Versc19}
{Verscharen}, D., {Klein}, K.~G., \& {Maruca}, B.~A. 2019, arXiv e-prints,
  arXiv:1902.03448

\bibitem[{{Wicks} {et~al.}(2011){Wicks}, {Horbury}, {Chen}, \&
  {Schekochihin}}]{Wicks11}
{Wicks}, R.~T., {Horbury}, T.~S., {Chen}, C.~H.~K., \& {Schekochihin}, A.~A.
  2011, Physical Review Letters, 106, 045001

\bibitem[{{Withbroe} \& {Noyes}(1977)}]{Withb77}
{Withbroe}, G.~L., \& {Noyes}, R.~W. 1977, \araa, 15, 363

\end{thebibliography}

\end{document}